\documentclass{amsart}
\usepackage{eurosym}

\theoremstyle{definition}

\theoremstyle{remark}

\numberwithin{equation}{section}
\begin{document}
\title{Lie Symmetry Analysis and Similarity Solutions for the \\ Jimbo - Miwa
Equation and Generalisations}
\author{Amlan K Halder}
\address{Department of Mathematics, Pondicherry University, Puducherry - 605014,
India}
\email{amlanhalder1@gmail.com}
\thanks{}
\author{A Paliathanasis}
\address{Institute for Systems Science, Durban University of Technology, PO
Box 1334, Durban 4000, Republic of South Africa}
\email{anpaliat@phys.uoa.gr}
\thanks{}
\author{Rajeswari Seshadri}
\address{Department of Mathematics, Pondicherry University,
Puducherry - 605014, India}
\email{seshadrirajeswari@gmail.com}
\thanks{}
\author{PGL Leach}
\address{School of Mathematics, Statistics and Computer Science, University
of KwaZulu-Natal, Durban, South Africa and}
\address{Institute for Systems Science, Durban University of Technology,
Durban, South Africa}
\email{leachp@ukzn.ac.za}
\thanks{}
\date{28:05:2019}

\begin{abstract}
We study the Jimbo - Miwa equation and two of its extended forms, as proposed
by Wazwaz et al, using Lie's group approach. Interestingly, the
travelling - wave solutions for all the three equations are similar. Moreover,
we obtain certain new reductions which are completely different for each
of the three equations. For example, for one of the extended forms of the
Jimbo - Miwa equation, the subsequent reductions leads to a second - order
equation with Hypergeometric solutions. In certain reductions, we obtain
simpler first - order and linearisable second - order equations, which helps us
to construct the analytic solution as a closed - form function. The variation in
the nonzero Lie brackets for each of the different forms of the Jimbo - Miwa
also presents a different perspective. Finally, singularity analysis is
applied in order to determine the integrability of the reduced equations and
of the different forms of the Jimbo - Miwa equation.
\end{abstract}

\maketitle

%\curraddr{�}

%    Remove any unused author tags.

%    author one information

%    author two information

%\curraddr{Department of Mathematics and Natural Sciences, Core Curriculum Program,
%Prince Mohammad Bin Fahd University, Al Khobar 31952, Kingdom of Saudi Arabia}

%    author three information

%\dedicatory{}

\vspace{1.5cc}

\section{Introduction}

The Jimbo - Miwa equation in $1+3$ space dimensions is a partial differential equation
(PDE) and along with its two different extended forms is our subject of study.
The equation \cite{jm4} was proposed in $1983$ and has gained considerable
attention in the past with respect to its study of integrability using
various standard techniques. It is well known that this equation is a member
of the KP hierarchy and possesses wider physical applications. Significant
works were conducted by Dorizzi et al\cite{jm3}, Wang et al\cite{jm12}, Wazwaz
et al \cite{jm5,jm11}, Cao et al\cite{jm2} and many others. Cao et al\cite{jm2}
also provides a comprehensive list of most of the work done with respect to
the algebraic structure of the equation.

The Jimbo - Miwa equation is defined to be
\begin{equation}
u_{xxxy}+3u_{y}u_{xx}+3u_{x}u_{xy}+2u_{ty}-3u_{xz}=0,  \label{1.1}
\end{equation}%
while the extended forms given by Wazwaz et al \cite{jm5} are,
\begin{equation}
u_{xxxy}+3u_{y}u_{xx}+3u_{x}u_{xy}+2u_{ty}-3(u_{xz}+u_{yz}+u_{zz})=0,
\label{1.2}
\end{equation}%
and
\begin{equation}
u_{xxxy}+3u_{y}u_{xx}+3u_{x}u_{xy}+2(u_{tx}+u_{ty}+u_{tz})-3u_{xz}=0.
\label{1.3}
\end{equation}

We study the determination of solutions for the three equations by
using the method of Lie point symmetries. Lie symmetries are a powerful tools
for the analysis of nonlinear differential equations. The main idea of
Lie symmetries is to determine the transformations which leave the given
differential equation invariant. Then, by using normal coordinates, the
solution of the differential equation can be written in terms of the
invariant functions for the Lie symmetry vector and in that way to reduce
the number of independent variables in case of PDEs, or the order of a
ordinary differential equation (ODE) \cite{kumei,ibra,olver}. Applications
of Lie symmetries can be found for instance in \cite%
{app6,app5,app8,app1,app3,app4,app7,app2,ap11} and references therein.

The application of Lie's theory for equations (\ref{1.1}), (\ref{1.2}) and (%
\ref{1.3}) reveals that equation (\ref{1.1}) admits six Lie point symmetries,
equation (\ref{1.2}) admits ten Lie point symmetries while equation\ (\ref%
{1.3}) is invariant under a six - dimensional group of point transformations.

The Lie symmetries are applied in order to determine similarity solutions
for the equations under our consideration. What is more interesting is that
the three equations of our study admit the same travelling-wave reduced
equations with slight variation among each other. Reductions with other
similarity variables provide different results for each equation. For
example, with respect to equation (\ref{1.1}), certain reductions lead to
homogenous solvable PDEs of less order or to second - order equations which
are maximally symmetric. Specifically, for equation (\ref{1.2}) certain
similarity variable reduces the equation to a second - order equation
possessing a Hypergeometric solution. Certain reductions of (%
\ref{1.2}) and (\ref{1.1}) also lead to a third - order or second - order
equation with zero Lie point symmetries. We analyse such equations using
singularity analysis. According to the authors' knowledge most of the results
obtained here are new and cannot be found in the literature. We took the aid of a
symbolic manipulation code developed by Dimas et al \cite{jm6,jm7,jm8}. The
paper also discuss the integrability of the three forms of Jimbo - Miwa
using singularity analysis. It is shown that equations (\ref{1.1}), (\ref%
{1.2}) and (\ref{1.3}) satisfy the test.

The paper is arranged as follows: In Section $2,$ the preliminaries of Lie point
symmetry analysis and singularity analysis are mentioned. In Section $3,$ and
its subsequent subsections Lie's point symmetry analysis of equation (\ref%
{1.1}) is discussed elaborately. In sections $4$ and $5,$ analysis of
equations (\ref{1.2}) and (\ref{1.3}) is discussed. Section $6,$ details the
singularity analysis of the second - order equation obtained by the subsequent
reductions of equation (\ref{1.1}). In Section $7,$ the singularity analysis
for the PDEs is presented. In the end, a brief conclusion and proper
references are mentioned.

\section{Preliminaries}

In this Section, we briefly discuss the mathematical tools that we apply in
this work to study the PDEs of our consideration. More specifically, we give
the basic definitions for the theory of Lie symmetries and singularity
analysis.

\subsection{Lie symmetries}

Let
\begin{equation}
F(\mathbf{x},u_{\alpha},u_{\beta},u_{\delta},u_{\alpha\beta},...)=0,  \label{1.01}
\end{equation}%
where $\mathbf{x}=(x_{\alpha},x_{\beta},x_{\delta})$ are the set of independent variables,
$u_{\alpha}=\frac{\partial {u}}{\partial {x_{\alpha}}},$ $u_{\beta}=\frac{\partial {u}}{%
\partial {x_{\beta}}},$ and $u_{\alpha\beta}=\frac{\partial {u}}{\partial {x_{\alpha}x_{\beta}}}%
,$ and the subsequent terms can be defined henceforth. Under an
infinitesimal point transformation,
\begin{eqnarray}
\bar{x_{\alpha}} &=&x_{\alpha}+\epsilon \xi ^{\alpha}(x_{\alpha},x_{\beta},x_{\delta},u)+\mathcal{O}(\epsilon ^{2}),
\notag  \label{1.02} \\
\bar{x_{\beta}} &=&x_{\beta}+\epsilon \xi ^{\beta}(x_{\alpha},x_{\beta},x_{\delta},u)+\mathcal{O}(\epsilon ^{2}),
\notag  \\
\bar{x_{\delta}} &=&x_{\delta}+\epsilon \xi ^{\delta}(x_{\alpha},x_{\beta},x_{\delta},u)+\mathcal{O}(\epsilon ^{2}),
\notag  \\
\bar{u} &=&u+\epsilon \eta (x_{\alpha},x_{\beta},x_{\delta},u)+\mathcal{O}(\epsilon ^{2}),
\end{eqnarray}%
equation (\ref{1.01}) is said to be invariant if and only if,
\begin{equation}
F(\mathbf{x},u)=F(\bar{\mathbf{x}},\bar{u}).  \label{1.03}
\end{equation}%
The transformations forms a symmetry group, say, G, the generator of
which can be defined as,
\begin{equation}
\Gamma =\eta (x_{\alpha},x_{\beta},x_{\delta},u)\partial _{u}+\xi ^{\alpha}(x_{\alpha},x_{\beta},x_{\delta},u)\partial _{x_{\alpha}}+\xi ^{\beta}(x_{\alpha},x_{\beta},x_{\delta},u)\partial _{x_{\beta}}+\xi ^{\delta}(x_{\alpha},x_{\beta},x_{\delta},u)\partial _{x_{\delta}}.
\label{1.04a}
\end{equation}%
Therefore $\Gamma $, which is the generator of the infinitesimal
transformations, can be considered as a Lie point symmetry of equation (\ref%
{1.01}). Now one can use equation (\ref{1.04a}) to discuss the reduction of
the corresponding PDEs using the characteristic functions, obtained by
solving the associated Lagrange's system which is,
\begin{equation*}
\frac{dx_{\alpha}}{\xi ^{\alpha}}=\frac{dx_{\beta}}{\xi ^{\beta}}=\frac{dx_{\delta}}{\xi ^{\delta}}=\frac{du}{\eta }.
\end{equation*}

\subsubsection{Singularity analysis.\newline
}

Our second method of investigation is known as singularity analysis which
evolved at about the same time as symmetry analysis, \textit{i.e.} towards the
end of the nineteenth century. In essence, it is the determination of the
existence of singularities in the dependent variable(s) in the complex plane
of the independent variable. Indeed, it is interesting to note how analysis
developed following the pioneering work of Cauchy in complex analysis.
Equations which satisfy the analysis are said to possess the Painlev\'{e}
Property. This is an interesting descriptor as the first known application
of singularity analysis was due to Sophie Kowalevski who used the analysis
to determine the third integrable case of the spinning top \cite{Kowalevski
88 a}\footnote{%
The first two cases were due to Euler and Lagrange in the seventeenth
century.}. The treatment of the Painlev\'{e} Property in the classic text of
E L Ince \cite{Ince 27 a} is very clear. Later works by Ramani, Grammaticos
and Bountis \cite{Ramani 89 a} and Michael Tabor \cite{Tabor 89 a} have kept
the idea of proving integrability through singularity analysis before the
public eye.

A major development is to be found in the works of Ablowitz, Ramani and
Segur \cite{Ablowitz 78 a, Ablowitz 80 a, Ablowitz 80 b} who introduced a
straightforward approach to determining whether a given differential
equation possesses the Painlev\'e Property. The approach has become known as
the ARS algorithm and is very simple in its concept although there are times
when its application faces some serious challenges. There are three steps:

\begin{enumerate}
\item Determine the existence or otherwise of a singularity by making a
substitution of $y = a(x-x_0)^p$ into the differential equation (for
simplicity we consider a single dependent variable).\\

This is called the leading - order term. The nature of the singularity is
indicated by the value of $p$ and its location is $x_{0}$. Originally, it was
considered to be a negative integer, hence, the method of polelike
expansions, but over the years it was accepted that $p$ could equally be a
fraction which could be positive or negative as differentiation of a
positive fractional power eventually leads to a negative power and so a
singularity. As a practical point, it must be borne in mind that a
fraction requires the complex plane to be divided into segments by branch - line cuts. A multitude of these is not good if numerical work has eventually
to be undertaken. The terms which contribute to the evaluation of $p$ are
called the dominant terms.

\item A differential equation of order greater than one needs additional
constants of integration and the idea is to construct a Laurent expansion
about the singularity.\\

The additional constants of integration enter the expansion at powers called
resonances and these are identified by making the substitution $y =
a(x-x_0)^p + m (x-x_0)^{p+s}$ into the dominant terms\footnote{%
Usually the letter $r$ is used, but we use $s$ in deference to the
pioneering work of Kowalevskaya.}. The result is a polynomial in $m$. A
given coefficient enters the expansion when its coefficient is $m$. The
coefficient is arbitrary, \textit{i.e.} to be determined by the initial
conditions, if $m$ is zero. The coefficient of $m$ is a polynomial in $s$
and the solution of polynomial equals zero gives the requisite values of $s$%
. One of the resonances must take the value $-1$. This value is associated
with the location of the moveable singularity.

\item The Laurent expansion is substituted into the complete equation with
the coefficients of the resonant terms bring arbitrary and the other
coefficients determined.

This can be a tedious process, but, if all goes well, the solution is
inferred to be analytic albeit possibly restricted in applicability by
branch cuts.
\end{enumerate}

Parallel procedures for partial differential equations can be found in \cite%
{Weiss 83 a, Weiss 83 b, Weiss 84 a, Weiss 84 b, Weiss 84 c, Weiss 85 a,
Weiss 85 b}.

\section{The Symmetry Analysis for equation (\protect\ref{1.1})}

For (\ref{1.1}), we compute the Lie - Point symmetries,
\begin{eqnarray}
\Gamma _{1a} &=&\partial _{t},  \notag  \label{2.1} \\
\Gamma _{2a} &=&t\partial _{t}-\frac{u}{3}\partial _{u}+\frac{x}{3}\partial
_{x}+\frac{2z}{3}\partial _{z},  \notag \\
\Gamma _{3a} &=&\partial _{z},  \notag \\
\Gamma _{4a} &=&t\partial _{x}+\frac{2x}{3}\partial _{u},  \notag \\
\Gamma _{5a} &=&c_{1}(z)\partial _{y}+\frac{3tc_{1}^{\prime }(z)}{4}\partial
_{x}+\left( \frac{2xc_{1}^{\prime }(z)}{4}-\frac{3tyc_{1}^{\prime \prime }(z)%
}{4}\right) \partial _{u},  \notag \\
\Gamma _{6a} &=&E_{0}(t,z)\partial _{u},  \notag
\end{eqnarray}%
where $c_{1}(z)$ and $E_{0}(t,z)$ are arbitrary functions.

The nonzero Lie brackets are,
\begin{equation*}
\begin{split}
\lbrack \Gamma _{1a},\Gamma _{2a}]& =\Gamma _{1a}, \\
\lbrack \Gamma _{1a},\Gamma _{4a}]& =\partial_{x}, \\
\end{split}
\qquad
\begin{split}
\lbrack \Gamma _{2a},\Gamma _{3a}]& =-\frac{2\Gamma_{3a}}{3}, \\
\lbrack \Gamma _{2a},\Gamma _{4a}]& =\frac{2\Gamma_{4a}}{3}. \\
\end{split}%
\end{equation*}

\textbf{Case 1:\newline
} To study the travelling - wave reductions, we follow a simple procedure,
by which we consider $m\Gamma _{3a}-c\Gamma _{1a}$, to reduce the equation (%
\ref{1.1}) to a new PDE of dimension $1+2$, where $m$ is the wave number and
$c$ denotes the frequency. The similarity variables for this reduction are,
\begin{eqnarray}
w &=&mz-ct,  \notag  \label{2.2} \\
u(t,x,y,z) &=&v(x,y,w).  \notag \\
&&
\end{eqnarray}%
The possible reduced PDE is
\begin{equation}
v_{xxxy}+3v_{x}v_{xy}+3v_{y}v_{xx}-2cv_{yw}-3mv_{xw}=0.  \label{2.3}
\end{equation}%
The Lie - Point symmetries are,
\begin{eqnarray}
\Gamma _{1b} &=&\partial _{w},  \notag  \label{2.4} \\
\Gamma _{2b} &=&v\partial _{v}-3w\partial _{w}-x\partial _{x}-y\partial _{y},
\notag \\
\Gamma _{3b} &=&\partial _{x},  \notag \\
\Gamma _{4b} &=&w\partial _{x}-\left( \frac{2cx}{3}+my\right)\partial _{v}
\notag \\
\Gamma _{5b} &=&\partial _{y}.  \notag
\end{eqnarray}%
\textbf{Case 1a:\newline
} The translation, with respect to the variables $x,w$ and $y$ is obtained
for the equation (\ref{2.3}).  Therefore, we use the operator $k\Gamma
_{3b}+l\Gamma _{5b}+\Gamma _{1b}$, where $k$ and $l$ are the wave numbers, to obtain
the reduced ode,
\begin{equation}
k^{3}lp^{\prime \prime \prime \prime
}(q)+6k^{2}lp^{\prime }(q)p^{\prime \prime}(q)-\left(2cl+3km\right)p^{\prime \prime }(q)=0,  \label{2.5}
\end{equation}%
where $q=kx+ly+w$ and $v(x,y,w)=p(q)$. This can be ascertained by the reader
that the equation, (\ref{2.5}), is the corresponding travelling - wave
reduction for the PDE, equation (\ref{1.1}). We compute the symmetries for
equation (\ref{2.5}).  They are,
\begin{eqnarray}
\Gamma _{1c} &=&\partial _{q},  \notag  \label{2.6} \\
\Gamma _{2c} &=&\partial _{p},  \notag \\
\Gamma _{2c} &=&q\partial _{q}+\left( \frac{mq}{kl}+\frac{2cq}{3k^{2}}%
-p\right) \partial _{p}.  \notag
\end{eqnarray}

\textbf{Case 1b:\newline
} We first look at the reduction using $\Gamma _{1c}$, namely,
\begin{equation}
g^{\prime \prime \prime }(h)=\frac{10g^{\prime }(h)g^{\prime \prime }(h)}{%
g(h)}-\frac{15g^{\prime 3}}{g(h)^{2}}+\left( \frac{(2cl+3km)g(h)^{2}}{%
k^{3}l}-\frac{6g(h)}{k}\right) g^{\prime }(h),  \label{2.7}
\end{equation}%
where $h=p(q)$ and $g(h)=\frac{1}{p^{\prime }(q)}$.  Equation (\ref{2.7}%
) possesses a lone symmetry, $\partial _{h}$. We use this to reduce (\ref{2.7}%
) to a second - order equation.
\begin{equation}
b^{\prime \prime }(a)=\frac{3b^{\prime 2}}{b(a)}+\frac{10b^{\prime }(a)}{a}%
+\left( \frac{6a}{k}-\frac{2a^{2}c}{k^{3}}-\frac{3a^{2}m}{k^{2}l}\right)
b(a)^{3}+\frac{15b(a)}{a^{2}},  \label{2.8}
\end{equation}%
where $a=g(h)$ and $b(a)=\frac{1}{g^{\prime }(h)}$.  Equation (\ref{2.8})
is maximally symmetric. Therefore, the solution of (\ref{1.1}), can be
obtained in accordance to equation (\ref{2.8}).\newline

\textbf{Case 1c:\newline
} Next, we use $\Gamma_{2c}$ to study the reduction of equation (\ref{2.5}).%
\newline
The reduced third - order equation is,
\begin{equation}  \label{2.9}
g^{\prime \prime \prime }(h) = \left(\frac{2cl+3km}{k^{3}l}-\frac{6g(h)}{k}%
\right)g^{\prime }(h),
\end{equation}
where $h=q$ and $g(h)=p^{\prime }(q)$. The reduced third - order equation has two
symmetries. They are $\partial_{h}$ and $h\partial_{h}+\frac{%
(2c-6gk^{2})l+3km}{3k^{2}l}\partial_{g}$. The reduction with respect to $%
\partial_{h}$ leads to the second - order equation
\begin{equation}  \label{2.10}
b^{\prime \prime }(a) = \frac{3b^{\prime 2}}{b(a)}+\frac{%
(6ak^{2}l-2cl-3km)b(a)^{3}}{k^{3}l},
\end{equation}
where $a=g(h)$ and $b(a)=\frac{1}{g^{\prime }(h)}$.  Equation (\ref{2.10}%
) possesses $8$ symmetries and hence is linearisable.

The reduction with respect to $h\partial _{h}+\frac{(2c-6gk^{2})l+3km}{%
3k^{2}l}\partial _{g}$ leads to the second - order equation,
\begin{equation}
b^{\prime \prime }(a)=\frac{3b^{\prime 2}}{b(a)}+9b^{\prime }(a)b(a)+\frac{(6a+26k)b(a)^{3}}{k}-\frac{%
(12a^{2}+24ak)b(a)^{4}}{k},  \label{2.11}
\end{equation}%
where
\begin{equation*}
a=\frac{(6g(h)k^{2}l-2cl-3km)h^{2}}{6k^{2}l}~\text{and}~b(a)=\frac{3k^{2}l}{%
h^{2}(3hg^{\prime}(h)k^{2}l+6g(h)k^{2}l-2cl-3km)}.
\end{equation*}%
\newline
Equation (\ref{2.11}) has zero Lie - Point symmetries. Singularity
analysis of this equation is treated in Section $6.$ The reduction with
respect to $\Gamma _{3c}$ leads to a third - order equation with zero Lie
point symmetries. We omit mentioning the equation here considering the high
nonlinearity of the third - order equation. The equation is under study and
we shall be discussing it in our subsequent work.

\subsection{Further reductions for the Jimbo - Miwa equation.\\}

In this subsection, we study the reductions of equation (\ref{1.1}) with
respect to $\Gamma _{2a}$. The similarity variables are,
\begin{eqnarray*}
w_{1} &=&\frac{x}{t^{\frac{1}{3}}}, \\
w_{2} &=&\frac{z}{t^{\frac{2}{3}}}, \\
u(t,x,y,z) &=&\frac{v(y,w_{1},w_{2})}{t^{\frac{1}{3}}}.
\end{eqnarray*}%
The reduced PDE of $1+2$ form is,
\begin{equation}
9v_{w_{1}w_{2}}+(2-9v_{w_{1}w_{1}})v_{y}+4w_{2}v_{yw_{2}}+2w_{1}v_{yw_{1}}-9v_{w_{1}}v_{yw_{1}}-3v_{yw_{1}w_{1}w_{1}}=0.
\label{3.01}
\end{equation}%
The symmetries of equation (\ref{3.01}) are,
\begin{eqnarray}
\Gamma _{1d} &=&y\partial _{y}+w_{2}\partial _{w_{2}},  \notag  \label{3.02}
\\
\Gamma _{2d} &=&\partial _{y},  \notag \\
\Gamma _{3d} &=&\partial _{w_{1}}+\frac{2w_{1}}{9}\partial _{v},  \notag \\
\Gamma _{4d} &=&2\sqrt{w_{2}}\partial _{w_{1}}-\frac{y}{\sqrt{w_{2}}}%
\partial _{v},  \notag \\
\Gamma _{5d} &=&c_{2}(w_{2})\partial _{v},  \notag
\end{eqnarray}%
where $c_{2}(w_{2})$ is an arbitrary function with respect to $w_{2}$.%
\newline
\textbf{Case 2:\newline
} We use $\Gamma _{1d}$ for the reduction.  The similarity
variables are \ $\frac{w_{2}}{y}=w_{3},v(y,w_{1},w_{2})=v_{1}(w_{1},w_{3}).~$%
The reduced PDE in $1+1$ dimensions is,
\begin{eqnarray}
0 &=&-4w_{3}^{2}{v_{1}}_{w_{3}w_{3}}+9{v_{1}}_{w_{1}w_{3}}-2w_{1}w_{3}{v_{1}}%
_{w_{1}w_{3}}+9w_{3}{v_{1}}_{w_{1}}{v_{1}}_{w_{1}w_{3}}  \notag  \label{3.03}
\\
&&+3w_{3}{v_{1}}_{w_{3}}(-2+3{v_{1}}_{w_{1}w_{1}})+3w_{3}{v_{1}}%
_{w_{1}w_{1}w_{1}w_{3}}.  \notag
\end{eqnarray}%
The Lie point symmetries are $\Gamma _{1e}=\partial _{v_{1}}$ and $\Gamma
_{2e}=w_{1}\partial _{v_{1}}+\frac{9}{2}\partial _{w_{1}}.$We use $\Gamma
_{2e}$ for reduction. The similarity variables are $v_{1}(w_{1},w_{3})=\frac{%
w_{1}^{2}}{9}+v_{2}(w_{3}).~$The reduced ode is $v_{2}^{\prime
}+w_{3}v_{2}^{\prime \prime }=0$ which  is maximally symmetric. Finally, the
closed - form similarity solution of equation (\ref{1.1}) is,
\begin{equation*}
u(t,x,y,z)=\frac{x^{2}}{9t}+\frac{K_{1}}{t^{\frac{1}{3}}}+\frac{K_{0}}{t^{%
\frac{1}{3}}}\log \left( {\frac{z}{yt^{\frac{2}{3}}}}\right) +\frac{K_{2}}{%
t^{\frac{1}{3}}}+\frac{K_{3}}{t^{\frac{1}{3}}},
\end{equation*}%
where $K_{0},$ $K_{1},$ $K_{2},$ and $K_{3}$ are arbitrary constants.\newline
\textbf{Case 3: Reduction with respect to $\Gamma _{2d}$:\newline
} The similarity variable is $v(y,w_{1},w_{2})=v_{1}(w_{1},w_{2})$. The
reduced PDE is~$~v_{{1}_{w_{1}w_{2}}}=0.$ The solution of equation (\ref{1.1}%
) can be given as,\newline
\begin{equation}
u(t,x,y,z)=\frac{f\left( \frac{x}{t^{\frac{1}{3}}}\right) +g\left( \frac{z}{%
t^{\frac{2}{3}}}\right) }{t^{\frac{1}{3}}}.  \label{3.006}
\end{equation}

\textbf{Case 4: Reduction with respect to $\Gamma _{3d}$:\newline
}
The similarity variable is,
\begin{equation*}
v(y,w_{1},w_{2})=\frac{w_{1}^{2}}{9}+v_{1}(y,w_{2}).
\end{equation*}%
The reduced PDE is $v_{{1}_{yw_{2}}}=0.~$ Similarly, to the above case,
the solution of the PDE can be easily determined. Therefore, the solution of the
equation (\ref{1.1}) can be given in terms of the solution of the reduced PDE.\newline
\textbf{Case 5: Reduction with respect to $\Gamma _{4d}$:\newline
} The similarity variables are $v(y,w_{1},w_{2})=-\frac{yw_{1}}{2w_{2}}%
+v_{1}(y,w_{2}).$ Hence, the reduced PDE is,
\begin{equation}
\frac{9y}{4w_{2}^{2}}+2v_{{1}_{y}}+4w_{2}v_{{1}_{yw_{2}}}=0.  \label{3.08}
\end{equation}%
The Lie - point symmetries for (\ref{3.08}) are derived to be,
\begin{eqnarray}
\Gamma _{1f} &=&c_{3}(y)\partial _{y},  \notag  \label{3.09} \\
\Gamma _{2f} &=&c_{4}(w_{2})\partial _{w_{2}}-\frac{v_{1}c_{4}(w_{2})}{2w_{2}%
}\partial _{v_{1}},  \notag \\
\Gamma _{3f} &=&v_{1}\partial _{v_{1}},  \notag \\
\Gamma _{4f} &=&E_{1}(y,w_{2})\partial _{v_{1}},  \notag
\end{eqnarray}%
where $c_{3}(y)$, $c_{4}(w_{2})$ and $E_{1}(y,w_{2})$ are arbitrary
functions. None of the above mentioned symmetries or linear combination of
any of them leads to a satisfactory reduction. The Painlev\'{e} analysis of
the equation (\ref{3.08}) is under study to ascertain its integrability.%
\newline
\textbf{Case 6: Reduction with respect to $\Gamma _{4a}$:\newline
} Next, we study the reduction with respect to $\Gamma _{4a}$ for equation (%
\ref{1.1}). The similarity variable is $u(t,x,y,z)=\frac{x^{2}}{3t}%
+v(t,y,z).~$The reduced PDE of dimension $1+2$ is,
\begin{equation}
v_{y}+tv_{ty}=0.  \label{3.010}
\end{equation}%
The reduced PDE possesses the following Lie point symmetries,
\begin{eqnarray*}
\Gamma _{1g} &=&E_{2}(t,z)\partial _{t}, \\
\Gamma _{2g} &=&E_{3}(y,z)\partial _{y}, \\
\Gamma _{3g} &=&E_{4}(z)\partial _{z}+vE_{4}(z)\partial _{v}, \\
\Gamma _{4g} &=&E_{5}(t,y,z)\partial _{v},
\end{eqnarray*}%
where $E_{i}$'s are arbitrary functions of the variables mentioned against
them. Equation (\ref{3.010}) can easily be integrated $v+tv_{t}+f\left(
t\right) =0$, $\ $from which we find $v\left( t\right) =\frac{v_{0}}{t}-%
\frac{\int f\left( t\right) dt}{t}.$

\subsection{Reductions for the Jimbo - Miwa equation.\newline
}

In this subsection, we study subsequent reductions for equations (\ref{2.3}%
) and (\ref{1.1}).  We start with $\Gamma _{2b}$, for which we consider
different possibilities for the similarity variables and study the
subsequent reductions. We consider firstly the similarity variables,
\begin{eqnarray}
\frac{x^{3}}{w} &=&w_{1},  \notag  \label{3.10} \\
\frac{x}{y} &=&w_{2},  \notag \\
v(x,y,w) &=&\frac{v_{1}(w_{1},w_{2})}{x}.  \notag
\end{eqnarray}%
The reduced PDE is,
\begin{eqnarray}
0 &=&-3w_{2}^{2}v_{1}(w_{1},w_{2})\left( 3w_{1}v_{{1}_{w_{1}w_{2}}}-2v_{{1}%
_{w_{2}}}+w_{2}v_{{1}_{w_{2}w_{2}}}\right) +2cw_{1}^{2}w_{2}^{2}v_{{1}%
_{w_{1}w_{2}}}-9mw_{1}^{3}v_{{1}_{w_{1}w_{1}}}  \notag  \label{3.11} \\
&&-3mw_{1}^{2}w_{2}v_{{1}_{w_{1}w_{2}}}-6mw_{1}^{2}v_{{1}%
_{w_{1}}}+27w_{1}^{3}w_{2}^{2}v_{{1}%
_{w_{1}w_{1}w_{1}w_{2}}}+27w_{1}^{2}w_{2}^{3}v_{{1}_{w_{1}w_{1}w_{2}w_{2}}}
\notag \\
&&+27w_{1}^{2}w_{2}^{2}v_{{1}_{w_{1}}}v_{{1}%
_{w_{1}w_{2}}}+54w_{1}^{2}w_{2}^{2}v_{{1}%
_{w_{1}w_{1}w_{2}}}+9w_{1}w_{2}^{4}v_{{1}%
_{w_{1}w_{2}w_{2}w_{2}}}+9w_{1}w_{2}^{3}v_{{1}_{w_{1}}}v_{{1}_{w_{2}w_{2}}}
\notag \\
&&+18w_{1}w_{2}^{3}v_{{1}_{w_{1}w_{2}w_{2}}}+6w_{1}w_{2}^{2}v_{{1}%
_{w_{1}w_{2}}}+3w_{2}^{2}v_{{1}_{w_{2}}}\left( 9w_{1}\left( w_{2}v_{{1}%
_{w_{1}w_{2}}}+w_{1}v_{{1}_{w_{1}w_{1}}}\right) +2w_{2}^{2}v_{{1}%
_{w_{2}w_{2}}}\right)   \notag \\
&&+w_{2}^{5}v_{{1}_{w_{2}w_{2}w_{2}w_{2}}}-6w_{2}^{3}v_{{1}_{w_{2}}}^{2}.
\notag
\end{eqnarray}%
The Lie-Point symmetries are,
\begin{eqnarray}
\Gamma _{1h} &=&\left( \frac{9mw_{1}}{4cw_{2}}-2w_{1}\right) \partial
_{w_{1}}+\left( \frac{3m}{4c}+w_{2}\right) \partial _{w_{2}}-\left( \frac{%
m^{2}w_{1}}{4cw_{2}^{2}}+\frac{3mv_{1}}{4cw_{2}}+\frac{cw_{1}}{9}+\frac{%
mw_{1}}{2w_{2}}-v_{1}\right) \partial _{v_{1}},  \notag  \label{3.12} \\
\Gamma _{2h} &=&w_{1}^{\frac{1}{3}}\partial _{v_{1}}.  \notag
\end{eqnarray}%
\textbf{Case 7:\newline
} We use $\Gamma _{2h}$, for reduction and the corresponding similarity
variables can be represented as $v_{1}(w_{1},w_{2})=p_{1}(q_{1},q_{2})$,
where $w_{1}=q_{1}$ and $w_{2}=q_{2}$. We assume that $%
p_{1}(q_{1},q_{2})=p_{1a}(q_{1})p_{2a}(q_{2})$, where $p_{1a}$ and $p_{2a}$
are arbitrary functions of $q_{1}$ and $q_{2}$ respectively. The computation
becomes less tedious for two particular cases, firstly when $p_{2a}(q_{2})=F_{0}
$, where $F_{0}$ is an arbitrary constant. The reduced second - order ode
which is maximally symmetric is,
\begin{equation}
2p_{1a}^{\prime }+3q_{1}(p_{1a}^{\prime \prime })=0.  \label{3.13}
\end{equation}%
Therefore, the solution of equation (\ref{1.1}) can be given in terms of
equation (\ref{3.13}).\newline
Similarly, when $p_{1a}(q_{1})=F_{1}$, where $F_{1}$ is arbitrary, we obtain
a fourth - order equation, namely,
\begin{equation}
6p_{2a}(q_{2})p_{2a}^{\prime }(q_{2})-6q_{2}{p_{2a}}^{\prime
}(q_{2})^{2}-3q_{2}p_{2a}(q_{2})p_{2a}^{\prime \prime
}(q_{2})+6q_{2}^{2}p_{2a}^{\prime }(q_{2})p_{2a}^{\prime \prime
}(q_{2})+q_{2}^{3}p_{2a}^{\prime \prime \prime \prime }(q_{2}) =0.
\label{3.14} \\
\end{equation}%
This equation has a single symmetry, $\partial_{q_{2}}$, which reduces equation (\ref{3.14}%
) to a third - order equation with zero point symmetries.  The equation is,
\begin{eqnarray}
p_{3a}^{\prime \prime \prime }(q_{3}) &=&\left( \frac{10p_{3a}^{\prime
}(q_{3})}{p_{3a}(q_{3})}+6p_{3a}(q_{3})\right) p_{3a}^{\prime \prime
}(q_{3})-15\frac{p_{3a}^{\prime }(q_{3})^{3}}{p_{3a}(q_{3})^{2}}%
-18p_{3a}^{\prime }(q_{3})^{2}  \notag  \label{3.15} \\
&&+\left( (3q_{3}-11)p_{3a}(q_{3})^{2}-6p_{3a}(q_{3})\right) p_{3a}^{\prime
}(q_{3})-12p_{3a}(q_{3})^{3}+(9q_{3}-6)p_{3a}(q_{3})^{4},  \notag \\
&&
\end{eqnarray}%
where $q_{3}=p_{2a}(q_{2})$ and $p_{3a}(q_{3})=\frac{1}{q_{2}p_{2a}^{\prime
}(q_{2})}.$ Singularity analysis of the equation (\ref{3.15}) is under study.%
\newline
\textbf{Case 8:\newline
} The similarity variable considered using $\Gamma _{4b}$ is $%
v(x,y,w)=-\frac{cx^{2}+3mxy}{3w}+v_{1}(y,w)$, where $y$ and $w$ are the new
independent variables. The reduced PDE is,
\begin{equation}
v_{{1}_{y}}+wv_{{1}_{yw}}=0.  \label{3.16}
\end{equation}%
The reduced PDE (\ref{3.16}) is similar to equation (\ref{3.010}) which
is also a reduced PDE of $1+1$ dimensions. Also, it can be easily observed
that equation(\ref{3.16}) is a variant of equation (\ref{3.08}). The Lie
point symmetries are almost in similar nature, hence we omit mentioning them
here.

\section{The Symmetry Analysis for equation (\protect\ref{1.2})}

The Lie-point symmetries are,
\begin{eqnarray}
\Gamma _{1i} &=&\partial _{x},  \notag  \label{4.1} \\
\Gamma _{2i} &=&\partial _{y},  \notag \\
\Gamma _{3i} &=&\frac{t}{2}\partial _{x}+\left( \frac{t}{2}+\frac{z}{3}%
\right) \partial _{y}+t\partial _{z},  \notag \\
\Gamma _{4i} &=&\frac{1}{2}\partial _{y}+\partial _{z},  \notag \\
\Gamma _{5i} &=&t\partial _{t}+\left( \frac{t}{2}+\frac{x}{3}+\frac{z}{12}%
\right) \partial _{x}+\left( \frac{3t}{8}+\frac{z}{4}\right) \partial _{y}+%
\frac{z}{2}\partial _{z}-\left( \frac{u}{3}+\frac{y}{12}\right) \partial
_{u},  \notag \\
\Gamma _{6i} &=&\partial _{t}+\frac{3}{8}\partial _{y},  \notag \\
\Gamma _{7i} &=&\frac{z}{2}\partial _{x}+\left( 2y-\frac{z}{2}-\frac{3t}{4}%
\right) \partial _{y}+z\partial _{z}-\frac{y}{2}\partial _{u},  \notag \\
\Gamma _{8i} &=&\frac{3c_{5}(t)}{2}\partial _{x}+xc_{5}^{\prime }(t)\partial
_{u},  \notag \\
\Gamma _{9i} &=&-\frac{3c_{6}(t)}{2}\partial _{x}+\left( zc_{6}^{\prime
}(t)-xc_{6}^{\prime }(t)\right) \partial _{u},  \notag \\
\Gamma _{10i} &=&c_{7}(t)\partial _{u},  \notag
\end{eqnarray}%
where $c_{5},$ $c_{6},$ and $c_{7}$ are arbitrary functions of $t.$ The
nonzero Lie brackets are,
\begin{equation*}
\begin{split}
\lbrack \Gamma _{1i},\Gamma _{5i}]& =\frac{\Gamma _{1i}}{3}, \\
\lbrack \Gamma _{2i},\Gamma _{5i}]& =-\frac{\partial_{u}}{12}, \\
\lbrack \Gamma _{2i},\Gamma _{7i}]& =-\frac{\partial_{u}}{2}+2\partial_{y}, \\
\lbrack \Gamma _{3i},\Gamma _{4i}]& =-\frac{\Gamma _{2i}}{3}, \\
\lbrack \Gamma _{3i},\Gamma _{5i}]& =-\left(\frac{3t+2z}{72}\right)\partial_{u}-\frac{t}{4%
}\partial_{x}-\left(\frac{t}{4}+\frac{z}{6}\right)\partial_{y}-\frac{t}{2}%
\partial_{z}, \\
\lbrack \Gamma _{3i},\Gamma _{6i}]& =-\frac{\Gamma _{1i}}{2}-\Gamma_{4i}, \\
\end{split}%
\qquad
\begin{split}
\lbrack \Gamma _{3i},\Gamma _{6i}]& =-\frac{\Gamma _{1i}}{2}-\Gamma_{4i}, \\
\lbrack \Gamma _{3i},\Gamma _{7i}]& =-\left(\frac{t}{4}+\frac{z}{6}%
\right)\partial_{u}+\frac{t}{2}\partial_{x}+\frac{3t+2z}{6}%
\partial_{y}+t\partial_{z}, \\
\lbrack \Gamma _{4i},\Gamma _{5i}]& =-\frac{\partial_{u}}{24}+\frac{%
\partial_{x}}{12}+\frac{\partial_{y}}{4}+\frac{\partial_{z}}{2}, \\
\lbrack \Gamma _{4i},\Gamma _{7i}]& =-\frac{\partial_{u}}{4}+\frac{%
\partial_{x}}{2}+\frac{\partial_{y}}{2}+\partial_{z}, \\
\lbrack \Gamma _{5i},\Gamma _{6i}]& =-\partial_{t}+\frac{\partial_{u}}{32}-%
\frac{\partial_{x}}{2}-\frac{3\partial_{y}}{8}, \\
\lbrack \Gamma _{5i},\Gamma _{7i}]& =-\left(\frac{t}{4}+\frac{z}{6}%
\right)\partial_{u}, \\
\lbrack \Gamma _{6i},\Gamma _{7i}]& = -\frac{3\partial_{u}}{16}. \\
\end{split}%
\end{equation*}

\textbf{Case 9:\newline
} We study the reduction firstly with the travelling - wave simplification. It
can be easily verified that $\Gamma _{4i}-\frac{\Gamma _{2i}}{2}$ and $%
\Gamma _{6i}-\frac{3\Gamma _{2i}}{8}$ are symmetries. Therefore, we use
linear combination of $\Gamma _{1i}$, $\Gamma _{2i}$, $\Gamma _{4i}-\frac{%
\Gamma _{2i}}{2}$ and $\Gamma _{6i}-\frac{3\Gamma _{2i}}{8}$, i.e. $k\Gamma
_{1i}+l\Gamma _{2i}+m(\Gamma _{4i}-\frac{\Gamma _{2i}}{2})-c(\Gamma _{6i}-%
\frac{3\Gamma _{2i}}{8})$, where $k,l,m$ are wave numbers and $c$ is the
frequency, to reduce the equation to a fourth-order ode,
\begin{equation}
k^{3}lp^{\prime \prime
\prime \prime }(q)+6 k^{2}lp^{\prime }(q)p^{\prime \prime }(q)-\left(2cl+3km+3lm+3m^{2}\right)p^{\prime \prime }(q)=0,  \label{4.2}
\end{equation}%
where $q=kx+ly+mz-ct$ and $p(q)=u(t,x,y,z)$. It is to be noted here that
equation (\ref{4.2}) is similar to the equation (\ref{2.5}). The
only difference can be found in the coefficient of second derivative of $p$
with respect to $q$. Also, the subsequent reductions are similar to the
previous section.\newpage

\subsection{Further Reductions of Equation (\protect\ref{1.2}).\newline
}

\textbf{Case 10:\newline
} We reduce equation (\ref{1.2}) using $\Gamma _{5i}$. It is to be
mentioned here that other symmetries such as, $\Gamma _{3i}$ and $\Gamma
_{7i}$ do not provide favourable reductions.\newline
The similarity variables for $\Gamma _{5i}$ are,
\begin{eqnarray*}
\frac{-3t+8y-4z}{8} &=&w_{1}, \\
\frac{z}{\sqrt{t}} &=&w_{2}, \\
\frac{4x-2z-3t}{4t^{\frac{1}{3}}} &=&w_{3}, \\
u(t,x,y,z) &=&\frac{45t^{\frac{4}{3}}-160t^{\frac{1}{3}}y+48t^{\frac{1}{3}%
}z+640v_{1}(w_{1},w_{2},w_{3})}{640t^{\frac{1}{3}}}.
\end{eqnarray*}%
The reduced PDE of dimension $1+2$ is
\begin{equation*}
9v_{{1}_{w_{2}w_{2}}}+(2-9v_{{1}_{w_{3}w_{3}}})v_{{1}_{w_{1}}}-2w_{3}v_{{1}_{w_{1}w_{3}}}-9v_{{1}%
_{w_{3}}}v_{{1}_{w_{1}w_{3}}}-3v_{{1}_{w_{1}w_{3}w_{3}w_{3}}}+3w_{2}v_{{1}_{w_{1}w_{2}}}=0.
\end{equation*}%
The Lie point symmetries are,
\begin{eqnarray}
\Gamma _{1j} &=&2w_{1}\partial _{w_{1}}+w_{2}\partial _{w_{2}},  \notag
\label{4.04} \\
\Gamma _{2j} &=&\partial _{w_{1}},  \notag \\
\Gamma _{3j} &=&\partial _{v_{1}},  \notag \\
\Gamma _{4j} &=&-\frac{9}{2}\partial _{w_{3}}+w_{3}\partial _{v_{1}},  \notag
\\
\Gamma _{5j} &=&w_{2}\partial _{v_{1}}.  \notag
\end{eqnarray}%
\textbf{Case 11:\newline
} We study the reduction with respect to $\Gamma _{1j}$. The similarity
variables are $w_{4}=\frac{w_{2}}{\sqrt{w_{1}}}%
~,~v_{1}(w_{1},w_{2},w_{3})=v_{2}(w_{3},w_{4}).$ The reduced PDE of $1+1$
dimensions is,
\begin{equation}
3(-6+w_{4}^{2})v_{{2}_{w_{4}w_{4}}}+w_{4}v_{{2}_{w_{4}}}(5-9v_{{2}%
_{w_{3}w_{3}}})-w_{4}\left( (2w_{3}+9v_{{2}_{w_{3}}})v_{{2}%
_{w_{3}w_{4}}}+3v_{{2}_{w_{3}w_{3}w_{3}w_{4}}}\right) =0.  \notag
\label{4.05}
\end{equation}%
The Lie point symmetries are $\Gamma _{1k}=\partial _{w_{3}}-\frac{2w_{3}}{9}%
\partial _{v_{2}},\Gamma _{2k}=\partial _{v_{2}}.$

\textbf{Case 12:\newline
} We study the reduction with $\Gamma _{1k}$. The similarity variable is,
\begin{equation*}
v_{2}(w_{3},w_{4})=-\frac{w_{3}^{2}}{9}+v_{3}(w_{4}).
\end{equation*}%
The reduced ode is,
\begin{equation}
7w_{4}v_{3}^{\prime }(w_{4})-18v_{3}^{\prime \prime
}(w_{4})+3w_{4}^{2}v_{3}^{\prime \prime }(w_{4})=0.  \label{4.07}
\end{equation}%
Equation (\ref{4.07}) is maximally symmetric. Also, one interesting
observation regarding (\ref{4.07}) is that the solution of the equation is
in terms of Hypergeometric function. Therefore, solution of (\ref{1.2}) can
be given in terms of equation (\ref{4.07}).\newline
\textbf{Case 13:\newline
} We study the reduction using $\Gamma _{2j}+\Gamma _{3j},$ the similarity
variables $w_{2},$ $w_{3},$ are the newly defined independent variable and $%
v_{1}(w_{1},w_{2},w_{3})=w_{1}+v_{2}(w_{2},w_{3}).$ The reduced PDE is,
\begin{equation}
2-9v_{{2}_{w_{3}w_{3}}}+9v_{{2}_{w_{2}w_{2}}}=0.  \label{4.08}
\end{equation}%
The symmetries of equation (\ref{4.08}) are,
\begin{eqnarray*}
\Gamma _{1l} &=&v_{2}\partial _{v_{2}}, \\
\Gamma _{2l} &=&\partial _{w_{3}}, \\
\Gamma _{3l} &=&\left( -c_{8}(w_{3}-w_{2})+c_{9}(w_{2}+w_{3})\right)
\partial _{w_{2}}+\left( -c_{8}(w_{3}-w_{2})+c_{9}(w_{2}+w_{3})\right)
\partial _{w_{3}}, \\
\Gamma _{4l} &=&E_{6}(w_{2},w_{3})\partial _{v_{2}},
\end{eqnarray*}%
where $c_{8}(w_{3}-w_{2}),c_{9}(w_{2}+w_{3})$ and $E_{6}(w_{2},w_{3})$ are
arbitrary functions.

\textbf{Case 14:\newline
} We use $\Gamma _{2l}$ for reduction. The similarity variable is $%
v_{2}(w_{2},w_{3})=v_{3}(w_{2}).$ The reduced ode is,
\begin{equation}
2+9v_{3}^{\prime \prime }(w_{2})=0.  \label{4.09}
\end{equation}%
Equation (\ref{4.09}) is linearisable. Therefore, the solution of (\ref{1.2}%
) can be given in terms of equation (\ref{4.09}).\newline
\textbf{Case 15:\newline}
We study the reduction with respect to $\Gamma _{4j}$. The
similarity variables are $v_{1}(w_{1},w_{2},w_{3})=-\frac{w_{3}^{2}}{9}%
+v_{2}(w_{1},w_{2}).$

The new independent variables are $w_{1}$ and $w_{2}$. The reduced PDE is,
\begin{equation}
9v_{{2}_{w_{2}w_{2}}}+4v_{{2}_{w_{1}}}+3w_{2}v_{{2}_{w_{1}w_{2}}}=0.
\label{4.10}
\end{equation}%
The Lie - point symmetries are,
\begin{eqnarray}
\Gamma _{1m} &=&\partial _{w_{1}},  \notag  \label{4.11} \\
\Gamma _{2m} &=&v_{2}\partial _{v_{2}},  \notag \\
\Gamma _{3m} &=&E_{7}(w_{1},w_{2})\partial _{v_{2}}.  \notag
\end{eqnarray}%
\textbf{Case 16:\newline
} The reduction using $\Gamma _{1m},$ leads to the well - known linearisable
second - order equation, which is\newline $\ v_{3}^{\prime \prime }(w_{2})=0,$ where
the similarity variable is $v_{2}(w_{1},w_{2})=v_{3}(w_{2}).$ Next, using $%
\Gamma _{1m}+\Gamma _{2m}$, which is a symmetry, the corresponding reduced
ode is,
\begin{equation}
4v_{3}(w_{2})+3w_{2}v_{3}^{\prime }(w_{2})+9v_{3}^{\prime \prime }(w_{2})=0,
\label{4.13}
\end{equation}%
where the similarity variables are $v_{2}(w_{1},w_{2})=e^{w_{1}}v_{3}(w_{2}).
$ \ Equation (\ref{4.13}) is a second - order linear equation and it is
maximally symmetric.

\section{The Symmetry Analysis for equation (\protect\ref{1.3})}

Similarly, for equation (\ref{1.3}), we only mention the Lie - point
symmetries here. The results with respect to the travelling - wave reduction
are similar and hence we omit it here.
\begin{eqnarray}
\Gamma _{1o} &=&\partial _{t},  \notag  \label{4.15} \\
\Gamma _{2o} &=&\partial _{x},  \notag \\
\Gamma _{3o} &=&\partial _{y},  \notag \\
\Gamma _{4o} &=&\partial _{z},  \notag \\
\Gamma _{5o} &=&f_{a}(t)\partial _{u},  \notag \\
\Gamma _{6o} &=&g_{a}(z)\partial _{u},  \notag
\end{eqnarray}%
where $f_{a}(t)$ and $g_{a}(z)$ are the arbitrary functions. It is clear
that the Lie brackets of the point symmetries, from $\Gamma_{1o}$ to $%
\Gamma_{6o},$ of equation (\ref{1.3}) do not possess any nonzero output and so the algebra is abelian .

\subsection{Further reduction for equation (\protect\ref{1.3}).\newline
}

\textbf{Case 17:\newline
} We study the reduction with respect to $\Gamma _{5o}+\Gamma _{1o}$. The
similarity variable is $u(t,x,y,z)=f_{1}(t)+v_{1}(x,y,z),$ where $%
f_{1}(t)=\int f_{a}(t)dt.$ The reduced PDE in $1+2$ dimensions is,
\begin{equation}
-3v_{{1}_{xz}}+3v_{{1}_{x}}v_{{1}_{xy}}+3v_{{1}_{y}}v_{{1}_{xx}}+v_{{1}%
_{xxxy}}=0.  \label{4.16}
\end{equation}%
The Lie - point symmetries are,
\begin{eqnarray}
\Gamma _{1p} &=&x\partial _{v_{1}}-z\partial _{y},  \notag  \label{4.17} \\
\Gamma _{2p} &=&\partial _{y},  \notag \\
\Gamma _{3p} &=&\partial _{z},  \notag \\
\Gamma _{4p} &=&h_{1}(z)\partial _{x}-yh_{1}^{\prime }(z)\partial _{v_{1}},
\notag \\
\Gamma _{5p} &=&h_{2}(z)\partial _{v_{1}},  \notag
\end{eqnarray}%
where $h_{1}(z)$ and $h_{2}(z)$ are arbitrary functions with respect to $z$.%
\newpage
\textbf{Case 18:\newline
} The similarity variable with respect to $\Gamma _{1p}$ is $v_{1}(x,y,z)=-%
\frac{xy}{z}+v_{2}(x,z).$ The reduced PDE in $1+1$ dimensions is,
\begin{equation}
v_{{2}_{x}}+zv_{{2}_{xz}}+xv_{{2}_{xx}}=0.  \label{3.001}
\end{equation}%
The Lie - point symmetries are,
\begin{eqnarray*}
\Gamma _{1q} &=&E_{8}(x,z)\partial _{x}+E_{8}(x,z)\partial _{v_{2}}, \\
\Gamma _{2q} &=&E_{9}(z)\partial _{z}, \\
\Gamma _{3q} &=&v_{2}E_{10}(z)\partial _{z},
\end{eqnarray*}%
where $E_{8}(x,z),E_{9}(z)$ and $E_{10}(z)$ are the arbitrary functions.%
\newline
The similarity variable for $\Gamma _{1q}$ is $v_{2}(x,z)=xv_{3}(z)$ and the
reduced ode is of Euler type, namely,
\begin{equation}
v_{3}(z)+zv_{3}^{\prime }(z)=0.  \label{3.002}
\end{equation}%
Next, the similarity variable using $\Gamma _{2q}+\Gamma _{3q},$ is of the
form $v_{2}(x,z)=e^{E_{11}(z)}+v_{3}(x),$ where $E_{11}(z)=\int \frac{%
E_{10}(z)}{E_{9}(z)}dz.$ The reduced ode is,
\begin{equation}
v_{3}^{\prime }(x)+xv_{3}^{\prime \prime }(x)=0.  \label{3.003}
\end{equation}%
Therefore, the solution of equation (\ref{1.3}) can be given in terms of
equation (\ref{3.002}) and (\ref{3.003}).\newline
\textbf{Case 19:\newline
} We use $\Gamma _{2p}+\Gamma _{3p}$ for the reduction. The similarity
variables are $w_{1}=z-y,~~v_{1}(x,y,z)=v_{2}(x,z-y).$ The reduced PDE is,
\begin{equation}
3v_{{2}_{xw_{1}}}+3v_{{2}_{x}}v_{{2}_{xw_{1}}}+3v_{{2}_{w_{1}}}v_{{2}%
_{xx}}+v_{{2}_{xxxw1}}=0.  \label{3.33}
\end{equation}%
The Lie - point symmetries of equation (\ref{3.33}) are,
\begin{eqnarray*}
\Gamma _{1r} &=&x\partial _{x}-\left( 2x+v_{2}\right) \partial _{v_{2}}, \\
\Gamma _{2r} &=&\partial _{x}, \\
\Gamma _{3r} &=&\partial _{v_{2}}, \\
\Gamma _{4r} &=&h_{3}(w_{1})\partial _{w_{1}},
\end{eqnarray*}%
where $h_{3}(w_{1})$ is an arbitrary function of $w_{1}.$\newline
We use $\Gamma _{1r}$ for reduction. The similarity variable, $%
v_{2}(x,w_{1})=-x+\frac{v_{3}(w_{1})}{x}$, leads to the reduced first - order
ode,
\begin{equation}
(-2+3v_{3}(w_{1}))v_{3}^{\prime }(w_{1})=0,  \label{3.34}
\end{equation}%
which can be solved easily.

\textbf{Case 20:\newline
} We now study the reduction with respect to $\Gamma _{6o}+\Gamma _{4o},$
for which the similarity variables are $u(t,x,y,z)=g_{1}(z)+v_{2}(t,x,y),$ and $%
g_{1}(z)=\int \frac{dz}{g_{a}(z)}.$ The reduced PDE of dimension $1+2$ is,
\begin{equation}
3v_{{2}_{x}}v_{{2}_{xy}}+3v_{{2}_{y}}v_{{2}_{xx}}+v_{{2}_{xxxy}}+2v_{{2}%
_{ty}}+2v_{{2}_{tx}}=0.  \label{3.341}
\end{equation}%
The Lie - point symmetries are,
\begin{eqnarray*}
\Gamma _{1s} &=&\partial _{t}, \\
\Gamma _{2s} &=&t\partial _{t}+\frac{x}{3}\partial _{x}+\frac{y}{3}\partial
_{y}-\frac{v_{2}}{3}\partial _{v_{2}}, \\
\Gamma _{3s} &=&\partial _{y}, \\
\Gamma _{4s} &=&t\partial _{x}+\left( \frac{2x}{3}+\frac{2y}{3}\right)
\partial _{v_{2}}, \\
\Gamma _{5s} &=&\partial _{x}, \\
\Gamma _{6s} &=&h_{4}(t)\partial _{v_{2}}.
\end{eqnarray*}%
where $h_{4}(t)$ is an arbitrary function of $t.$ We use $\Gamma
_{1s}+\Gamma _{3s}+\Gamma _{5s}$ to study the further reduction.

The similarity variables are,
\begin{eqnarray*}
x-t &=&w_{1}, \\
y-t &=&w_{2}, \\
v_{2}(t,x,y) &=&v_{3}(w_{1},w_{2}).
\end{eqnarray*}%
The reduced PDE of dimension $1+1$ is,
\begin{equation}
-2v_{{3}_{w_{2}w_{2}}}-4v_{{3}_{w_{1}w_{2}}}+3v_{{3}_{w_{1}}}v_{{3}%
_{w_{1}w_{2}}}-2v_{{3}_{w_{1}w_{1}}}+3v_{{3}_{w_{2}}}v_{{3}_{w_{1}w_{1}}}+v_{%
{3}_{w_{1}w_{1}w_{1}w_{2}}}=0.  \notag  \label{3.342}
\end{equation}%
The Lie - point symmetries are,
\begin{eqnarray*}
\Gamma _{1t} &=&\partial _{w_{1}}, \\
\Gamma _{2t} &=&\partial _{w_{2}}, \\
\Gamma _{3t} &=&\partial _{v_{3}}.
\end{eqnarray*}%
The similarity variables with respect to $\Gamma _{1t}+\Gamma _{2t}+\Gamma
_{3t}$ are,
\begin{eqnarray*}
w_{2}-w_{1} &=&w_{3}, \\
v_{3}(w_{1},w_{2}) &=&w_{1}+v_{4}(w_{3}). \\
&&
\end{eqnarray*}%
The reduced ode is,
\begin{equation}
v_{4}^{\prime \prime \prime \prime }(w_{3})-6v_{4}^{\prime }(w_{3})v_{4}^{\prime \prime
}(w_{3})+3v_{4}^{\prime \prime }(w_{3})=0.  \label{3.343}
\end{equation}%
Equation (\ref{3.343}) has three symmetries, which are,
\begin{eqnarray*}
\Gamma _{1u} &=&\partial _{w_{3}}, \\
\Gamma _{2u} &=&\partial _{v_{4}}, \\
\Gamma _{3u} &=&w_{3}\partial _{w_{3}}+\left( w_{3}-v_{4}\right) \partial
_{v_{4}}.
\end{eqnarray*}%
\textbf{Case 21:\newline
} The reduction with respect to $\Gamma _{1u},$ leads to a third - order ode,
\begin{equation}
v_{5}^{\prime \prime \prime }(w4)=\frac{10v_{5}^{\prime }(w4)v_{5}^{\prime
\prime }(w4)}{v_{5}(w4)}-\frac{15v_{5}^{\prime 3}}{v_{5}(w4)^{2}}%
+(-3v_{5}(w4)^{2}+6v_{5}(w4))v_{5}^{\prime }(w4),  \label{3.344}
\end{equation}%
where $w_{4}=v_{4}(w_{3})$ and $v_{5}(w_{4})=\frac{1}{v_{4}^{\prime }(w_{3})}%
.$ The equation (\ref{3.344}) has a sole symmetry, $\partial _{w_{4}}$. The
reduced third - order ode is further reduced to the second - order equation,
which is,
\begin{equation}
v_{6}^{\prime \prime }(w5)=\frac{3v_{6}^{\prime }(w_{5})^{2}}{v_{6}(w_{5})}+%
\frac{10v_{6}^{\prime }(w_{5})}{w_{5}}+\frac{%
(3w_{5}^{4}-6w_{5}^{3})v_{6}(w_{5})^{3}}{w_{5}^{2}}+\frac{15v_{6}(w5)}{%
w_{5}^{2}},  \label{3.345a}
\end{equation}%
where $w_{5}=v_{5}(w_{4})$ and $v_{6}(w_{5})=\frac{1}{v_{5}^{\prime }(w_{4})}%
.$ The equation (\ref{3.345a}) is maximally symmetric.\newline
\textbf{Case 22:\newline
} The reduction with respect to $\Gamma _{2u},$ leads to the third - order
equation,
\begin{equation}
v_{5}^{\prime \prime \prime }(w4)=(6v_{5}(w4)-3)v_{5}^{\prime }(w4),
\label{3.346a}
\end{equation}%
where $w_{4}=w_{3}$ and $v_{5}(w_{4})=v_{4}^{\prime }(w_{3}).$  Equation (%
\ref{3.346a}) has two symmetries, which are $\partial _{w_{4}}$ and $%
w_{4}\partial _{w_{4}}+\left( 1-2v_{5}\right) \partial _{v_{5}}.$ The
reduction with respect to $\partial _{w_{4}}$ leads to the second - order
equation,
\begin{equation}
v_{6}^{\prime \prime }(w5)=\frac{3v_{6}^{\prime 2}}{v_{6}(w5)}%
+(-6w_{5}+3)v_{6}(w_{5})^{3},  \label{3.347a}
\end{equation}%
where $w_{5}=v_{5}(w_{4})$ and $v_{6}(w_{5})=\frac{1}{v_{5}^{\prime }(w_{4})}%
.$ The equation (\ref{3.347a}) possess $8$ symmetries and hence is maximally
symmetric. Next, we use $w_{4}\partial _{w_{4}}+\left( 1-2v_{5}\right)
\partial _{v_{5}},$ for reduction. The reduced second - order ode is,
\begin{equation}
v_{6}^{\prime \prime }(w5)=\frac{3v_{6}^{\prime }(w_{5})^{2}}{v_{6}(w_{5})}%
+9v_{6}(w_{5})v_{6}^{\prime
}(w_{5})+(26-6w_{5})v_{6}(w_{5})^{3}+(12w_{5}^{2}-24w_{5})v_{6}(w_{5})^{4},
\label{3.348a}
\end{equation}%
where $w_{5}=\frac{(2v_{5}(w_{4})-1)w_{4}^{2}}{2}$ and $v_{6}(w_{5})=\frac{1%
}{w_{4}^{2}(w_{4}v_{5}^{\prime }(w_{4})+2v_{5}(w_{4})-1)}.$  Equation (%
\ref{3.348a}) possesses zero Lie point symmetries and it is similar to
equation (\ref{2.11}). The singularity analysis fails to infer any specific
conclusion for equation (\ref{3.348a}). Therefore, the solution of equation (%
\ref{1.3}) can be given in terms of equation (\ref{3.345a}) and (\ref{3.347a}%
).\newline
\textbf{Case 23:\newline
} The reduction with respect to $\Gamma _{3u}$ leads to a third - order
equation with zero Lie point symmetries. The equation is,
\begin{eqnarray}
v_{5}^{\prime \prime \prime }(w4) &=&\left(\frac{10v_{5}^{\prime }(w_{4})}{%
v_{5}(w_{4})}+10v_{5}(w_{4})\right)v_{5}^{\prime \prime }(w_{4})-\frac{%
15v_{5}^{\prime }(w_{4})^{3}}{v_{5}(w_{4})^{2}}-30v_{5}^{\prime }(w_{4})^{2}
\notag  \label{3.349} \\
&&+\left((-6w_{4}-35)v_{5}(w_{4})^{2}+6v_{5}(w_{4})\right)v_{5}^{\prime
}(w_{4})+18v_{5}(w_{4})^{3}-(30w_{4}+50)v_{5}(w_{4})^{4}
\notag \\
&&+(12w_{4}^{2}+24w_{4})v_{5}(w_{4})^{5},  \notag
\end{eqnarray}%
where $w_{4}=\frac{(2v_{4}(w_{3})-w_{3})w_{3}}{2}$ and $v_{5}(w_{4})=\frac{1%
}{w_{3}(w_{3}v_{4}^{\prime }(w_{3})+v_{4}(w_{3})-w_{3})}.$ Singularity
analysis of this equation is under study.\newline
\textbf{Case 24:\newline
} The reduction with respect to $\Gamma _{2s},$ leads to a PDE of dimension $%
1+1$, with similarity variable, $$v_{2}(t,x,y)=\frac{v_{3}(\frac{x}{t^{1}{3}},%
\frac{y}{t^{1}{3}})}{t^{\frac{1}{3}}},$$%
\begin{eqnarray}
0 &=&2w_{2}v_{{3}_{w_{2}w_{2}}}+4v_{{3}_{w_{1}}}+2w_{1}v_{{3}%
_{w_{1}w_{2}}}+2w_{2}v_{{3}_{w_{1}w_{2}}}-9v_{{3}_{w_{1}}}v_{{3}%
_{w_{1}w_{2}}}+v_{{3}_{w_{2}}}(4-9v_{{3}_{w_{1}w_{1}}})  \notag
\label{3.350a} \\
&&+2w_{1}v_{{3}_{w_{1}w_{1}}}-3v_{{3}_{w_{1}w_{1}w_{1}w_{2}}}.  \notag \\
&&
\end{eqnarray}%
Equation (\ref{3.350a}) has a sole point symmetry, $\partial _{v_{3}}.$
The subsequent reductions do not yield any favourable reductions.\newline
\textbf{Case 25:\newline
} The reductions with respect to $\Gamma _{4s},$ leads to a PDE of $1+1$
dimension. Similar equations are obtained in section $3,$ equation (\ref%
{3.010}), where we have discussed it in detail. The equation is,
\begin{equation}
v_{{3}_{y}}+tv_{{3}_{ty}}=0,  \label{3.351}
\end{equation}%
where the similarity variable is $v_{2}(t,x,y)=\frac{2(\frac{x^{2}}{2}+xy)}{%
3t}+v_{3}(t,y).~$

\section{Singularity analysis}

We study the singularity analysis of equation (\ref{2.11}). We suggest that
readers refer \cite{jm9,jm10,jm13,jm12,jm11} to understand the
preliminaries.\newline
Firstly, we write it in a convenient format,
\begin{equation}
-2(13k+3x)y(x)^{4}+12x(2k+x)y(x)^{5}-9ky(x)^{2}y^{\prime }(x)-3ky^{\prime
}(x)^{2}+ky(x)y^{\prime \prime }(x)=0,  \label{6.01}
\end{equation}%
where we mention $b(a)=y(x)$. We substitute $y\rightarrow \alpha x^{p}$ in
equation (\ref{6.01}) and look for the possible values of the exponent $p$.
The substitution leads to,
\begin{equation}
-26a^{4}kx^{4p}-a^{2}kpx^{-2+2p}-2a^{2}kp^{2}x^{-2+2p}-9a^{3}kpx^{-1+3p}-6a^{4}x^{1+4p}+24a^{5}kx^{1+5p}+12a^{5}x^{2+5p}=0
\label{6.02}
\end{equation}%
One of the possible values obtained from the dominant terms is $-1.$ For $%
p=-1$, the possible values of the leading-order coefficient $\alpha $ are $0,%
\frac{1}{4},\frac{1}{3},\frac{1}{2}.$ Next we substitute $y\rightarrow
\alpha x^{-1}+mx^{-1+s}$ into (\ref{6.01}) to compute the resonances ($s$
denotes the resonance). The substitution leads to,
\begin{eqnarray}
0 &=&-\frac{a^{2}k}{x^{4}}+\frac{9a^{3}k}{x^{4}}-\frac{26a^{4}k}{x^{4}}+%
\frac{24a^{5}k}{x^{4}}-\frac{6a^{4}}{x^{3}}+\frac{12a^{5}}{x^{3}}%
-2akmx^{-4+s}+27a^{2}kmx^{-4+s}-104a^{3}kmx^{-4+s}+  \notag  \label{6.03} \\
&&120a^{4}kmx^{-4+s}+3akmsx^{-4+s}-9a^{2}kmsx^{-4+s}+akms^{2}x^{-4+s}-24a^{3}mx^{-3+s}+60a^{4}mx^{-3+s}
\notag \\
&&-km^{2}x^{-4+2s}+27akm^{2}x^{-4+2s}-156a^{2}km^{2}x^{-4+2s}+240a^{3}km^{2}x^{-4+2s}+3km^{2}sx^{-4+2s}
\notag \\
&&-18akm^{2}sx^{-4+2s}-2km^{2}s^{2}x^{-4+2s}-36a^{2}m^{2}x^{-3+2s}+120a^{3}m^{2}x^{-3+2s}
\notag \\
&&+9km^{3}x^{-4+3s}-104akm^{3}x^{-4+3s}+240a^{2}km^{3}x^{-4+3s}-9km^{3}sx^{-4+3s}-24am^{3}x^{-3+3s}
\notag \\
&&+120a^{2}m^{3}x^{-3+3s}-26km^{4}x^{-4+4s}+120akm^{4}x^{-4+4s}-  \notag \\
&&6m^{4}x^{-3+4s}+60am^{4}x^{-3+4s}+24km^{5}x^{-4+5s}+12m^{5}x^{-3+5s}.
\end{eqnarray}%
We consider the linear terms with respect to $m$ from equation (\ref{6.03}),
\begin{eqnarray}
&&-2akx^{-4+s}+27a^{2}kx^{-4+s}-104a^{3}kx^{-4+s}+120a^{4}kx^{-4+s}+3aksx^{-4+s}-9a^{2}ksx^{-4+s}+
\notag  \label{6.04} \\
&&aks^{2}x^{-4+s}-24a^{3}x^{-3+s}+60a^{4}x^{-3+s}.
\end{eqnarray}%
We list the corresponding values of resonances for various nonzero values of
the leading - order coefficient, $\alpha $. $\left( \alpha =\frac{1}{2}%
\rightarrow s=(\frac{1}{2},1)\right) $, $\left( \alpha =\frac{1}{3}%
\rightarrow s=(-\frac{1}{3},\frac{1}{3})\right) $, $\left( \alpha =\frac{1}{4%
}\rightarrow s=(-\frac{1}{2},-\frac{1}{4})\right) $. It is to be observed
that the generic value of $s$, which is $-1$, is not obtained for any of the
possible values of the leading - order coefficient\cite{jm9,jm10} and hence
we cannot infer about the integrability of equation (\ref{2.11}).

\section{Singularity analysis for the Jimbo-Miwa PDE}

We apply the singularity analysis for the PDE (\ref{1.1}). We do that by
replacing $u\rightarrow U_{0}\phi (t,x,y,z)^{p}~$\cite{ptpde1,ptpde2}$,$ in
equation (\ref{1.1}).
\begin{eqnarray}
0 &=&3pU_{0}\Phi ^{-2+p}\Phi _{z}\Phi _{x}-3p^{2}U_{0}\Phi ^{-2+p}\Phi
_{z}\Phi _{x}-6p^{2}U_{0}^{2}\Phi ^{-3+2p}\Phi _{y}\Phi _{x}^{2}  \notag
\label{7.01} \\
&&+6p^{3}U_{0}^{2}\Phi ^{-3+2p}\Phi _{y}\Phi _{x}^{2}-6pU_{0}\Phi
^{-4+p}\Phi _{y}\Phi _{x}^{3}+  \notag \\
&&+11p^{2}U_{0}\Phi ^{-4+p}\Phi _{y}\Phi _{x}^{3}-6p^{3}U_{0}\Phi
^{-4+p}\Phi _{y}\Phi _{x}^{3}+  \notag \\
&&p^{4}U_{0}\Phi ^{-4+p}\Phi _{y}\Phi _{x}^{3}-3pU_{0}\Phi ^{-1+p}\Phi
_{xz}+3p^{2}U_{0}^{2}\Phi ^{-2+2p}\Phi _{x}\Phi _{xy}+  \notag \\
&&+6pU_{0}\Phi ^{-3+p}\Phi _{x}^{2}\Phi _{xy}-9p^{2}U_{0}\Phi ^{-3+p}\Phi
_{x}^{2}\Phi _{xy}+  \notag \\
&&+3p^{3}U_{0}\Phi ^{-3+p}\Phi _{x}^{2}\Phi _{xy}+3p^{2}U_{0}^{2}\Phi
^{-2+2p}\Phi _{y}\Phi _{xx}  \notag \\
&&+6pU_{0}\Phi ^{-3+p}\Phi _{y}\Phi _{x}\Phi _{xx}-9p^{2}U_{0}\Phi
^{-3+p}\Phi _{y}\Phi _{x}\Phi _{xx}+  \notag \\
&&+3p^{3}U_{0}\Phi ^{-3+p}\Phi _{y}\Phi _{x}\Phi _{xx}-3pU_{0}\Phi
^{-2+p}\Phi _{xy}\Phi _{xx}  \notag \\
&&+3p^{2}U_{0}\Phi ^{-2+p}\Phi _{xy}\Phi _{xx}-3pU_{0}\Phi ^{-2+p}\Phi
_{x}\Phi _{xxy}+3p^{2}U_{0}\Phi ^{-2+p}\Phi _{x}\Phi _{xxy}  \notag \\
&&-pU_{0}\Phi ^{-2+p}\Phi _{y}\Phi _{xxx}+p^{2}U_{0}\Phi ^{-2+p}\Phi
_{y}\Phi _{xxx}+pU_{0}\Phi ^{-1+p}\Phi _{xxxy}+  \notag \\
&&-2pU_{0}\Phi ^{-2+p}\Phi _{y}\Phi _{t}+2p^{2}U_{0}\Phi ^{-2+p}\Phi
_{y}\Phi _{t}+2pU_{0}\Phi ^{-1+p}\Phi _{ty}.
\end{eqnarray}%
Solving the dominant terms, we obtain the value of $p$ is $-1.$ To obtain
the coefficient of the leading - order exponent we substitute the value of $p$
and collect the dominant terms, which gives,
\begin{equation}
-12U_{0}^{2}\Phi _{y}\Phi _{x}^{2}+24U_{0}\Phi _{y}\Phi _{x}^{3}.
\label{7.02}
\end{equation}%
We solve equation (\ref{7.02}) to obtain the values of $U_{0},$ which are,
\begin{equation*}
U_{0}\rightarrow 0,U_{0}\rightarrow 2\phi _{x}.
\end{equation*}%
For the nonzero $U_{0},$ we compute the resonances, for which we substitute,
\begin{equation*}
U(t,x,y,z)\rightarrow U_{0}\Phi (t,x,y,z)^{-1}+m\Phi (t,x,y,z)^{-1+S}
\end{equation*}%
into equation (\ref{1.1}). The substitution leads to,
\begin{eqnarray*}
0 &=&6m^{2}(-2+S)(-1+S)^{2}\Phi ^{2S}\Phi _{y}\Phi _{x}^{2}-12U_{0}\Phi
_{y}(U_{0}-2\Phi _{x})\Phi _{x}^{2} \\
&&+m(-4+S)(-1+S)\Phi ^{S}\Phi _{y}\Phi _{x}^{2}(-6U_{0}+(S-3)(-2+S)\Phi
_{x})+ \\
&&+3m^{2}(-1+S)^{2}\Phi ^{1+2S}(\Phi _{x}\Phi _{xy}+\Phi _{y}\Phi _{xx})+ \\
&&+3U_{0}\Phi (U0-6\Phi _{x})(\Phi _{x}\Phi _{xy}+\Phi _{y}\Phi _{xx})+ \\
&&+3m(-1+S)\Phi ^{(}1+S)(-2U_{0}+(S-3)(-2+S)\Phi _{x})(\Phi _{x}\Phi
_{xy}+\Phi _{y}\Phi _{xx}) \\
&&-m(-2+S)(-1+S)\Phi ^{2+S}(-3\Phi _{xy}\Phi _{xx}+3\Phi _{x}(\Phi _{z}-\Phi
_{xxy})-\Phi _{y}(\Phi _{xxx}+2\Phi _{t})) \\
&&+2U_{0}\Phi ^{2}(3\Phi _{xy}\Phi _{xx}+3\Phi _{x}(-\Phi _{z}+\Phi
_{xxy})+\Phi _{y}(\Phi _{xxx}+2\Phi _{t}))+ \\
&&+U_{0}\Phi ^{3}(3\Phi _{xz}-\Phi _{xxxy}-2\Phi _{ty})-m(S-1)\Phi
^{3+S}(3\Phi _{x}-\Phi _{xxxy}-2\Phi _{ty}).
\end{eqnarray*}%
Next, we collect all the linear terms with respect to $m,$ which gives,
\begin{eqnarray}
&&(-4+S)(-1+S)\Phi ^{S}\Phi _{y}\Phi _{x}^{2}(-6U_{0}+(S-3)(-2+S)\Phi _{x})+
\notag  \label{7.04} \\
&&3(-1+S)\Phi ^{1+S}(-2U_{0}+(S-3)(-2+S)\Phi _{x})(\Phi _{x}\Phi _{xy}+\Phi
_{y}\Phi _{xx})  \notag \\
&&-(S-2)(-1+S)\Phi ^{2+S}(-3\Phi _{xy}\Phi _{xx}+3\Phi _{x}(\Phi _{z}-\Phi
_{xxy})  \notag \\
&&-\Phi _{y}(\Phi _{xxx}+2\Phi _{t}))-(S-1)\Phi ^{3+S}(3\Phi _{xz}-\Phi
_{xxxy}-2\Phi _{ty}).  \notag
\end{eqnarray}%
After substitution of the nonzero value of $U,$ we obtain,
\begin{equation}
-24\Phi _{y}\Phi _{x}^{3}+10S\Phi _{y}\Phi _{x}^{3}+23S^{2}\Phi _{y}\Phi
_{x}^{3}-10S^{3}\Phi _{y}\Phi _{x}^{3}+S^{4}\Phi _{y}\Phi _{x}^{3}.
\label{7.05}
\end{equation}%
When we factor equation (\ref{7.05}), we get the values of resonances as $%
s\rightarrow -1,1,4,6.$  As the generic values of $-1$ for $S$ is
obtained, we next verify the consistency, for which we substitute,
\begin{eqnarray*}
U &=&(2\Phi _{x})\Phi (t,x,y,z)^{-1}+U_{1}(t,x,y,z)\Phi ^{-1+1}+ \\
&&+U_{2}(t,x,y,z)\Phi (t,x,y,z)^{-1+2}+U_{3}(t,x,y,z)\Phi (t,x,y,z)^{-1+3} \\
&&+U_{4}(t,x,y,z)\Phi (t,x,y,z)^{-1+4}+ \\
&&+U_{5}(t,x,y,z)\Phi (t,x,y,z)^{-1+5}+U_{6}(t,x,y,z)\Phi (t,x,y,z)^{-1+6},
\end{eqnarray*}%
into equation (\ref{1.1}). From our observation, the values for $U_{2}(t,x,y,z)
$ and $U_{3}(t,x,y,z)$ are,
\begin{eqnarray*}
6\Phi _{y}\Phi _{x}^{2}U_{2}(t,x,y,z) &=&3\Phi _{z}\Phi _{x}-3\Phi _{y}U_{{1}%
_{x}}\Phi _{x}-3U_{{1}_{y}}\Phi _{x}^{2}+ \\
&&+3\Phi _{xy}\Phi _{xx}-3\Phi _{x}\Phi _{xxy}-\Phi _{y}\Phi _{xxx}-2\Phi
_{y}\Phi _{x}
\end{eqnarray*}%
\\
\begin{equation*}
U_{3}(t,x,y,z)=\frac{1}{12\Phi _{y}\Phi _{x}^{3}}\left(
\begin{array}{c}
3\Phi _{y}U_{{2}_{x}}\Phi _{x}^{2}-3U_{{2}_{y}}\Phi _{x}^{3}-6\Phi _{x}\Phi
_{xz}+3\Phi _{x}^{2}U_{{1}_{xy}}+ \\
+6U_{{1}_{x}}\Phi _{x}\Phi _{xy}+9U_{2}\Phi _{x}^{2}\Phi _{xy}+3\Phi
_{y}\Phi _{x}U_{{1}_{xx}}-3\Phi _{z}\Phi _{xx}+ \\
+3\Phi _{y}U_{{1}_{x}}\Phi _{xx}+9U_{{1}_{y}}\Phi _{x}\Phi _{xx}+15U_{2}\Phi
_{y}\Phi _{x}\Phi _{xx} \\
-2\Phi _{xy}\Phi _{xxx}+4\Phi _{x}\Phi _{xxxy}+\Phi _{y}\Phi _{xxxx}+2\Phi
_{xy}\Phi _{t}+2\Phi _{x}\Phi _{ty}+2\Phi _{y}\Phi _{tx}.%
\end{array}%
\right)
\end{equation*}%
As, observed from the values of $U_{2}(t,x,y,z)$ and $U_{3}(t,x,y,z),$
arbitrary term is $U_{1}(t,x,y,z).$ Similar observations can be made from $%
U_{4}(t,x,y,z)$ to $U_{6}(t,x,y,z)$.  The calculations are bit tedious and we
omit mentioning them here. We summarize our analysis in the following theorem.

\textbf{Theorem:} The Jimbo-Miwa equation (\ref{1.1}) and its generalizations
(\ref{1.2}) and (\ref{1.3}) pass the singularity test, that is, have the
Painlev\'{e} property and are integrable. Their solution is expressed by a
Right Painlev\'{e} Series. \newline

\section{Conclusion}

An elaborate study of the reductions of Jimbo - Miwa and its extended forms is
discussed. As mentioned above, new reductions of the equations were
obtained. The singularity analysis of certain reduced odes and all the three
forms of Jimbo - Miwa equation are discussed. This paper forms the first part
of our subsequent papers on Jimbo - Miwa equations. The conservation laws and
the solutions from them are under study. Moreover, we have discussed
reductions elaborately, but not exhaustively. Therefore, the
remaining reductions will also be discussed in our future work.

Finally, by applying the singularity analysis for the PDEs of our
consideration, without applying any similarity transformation, we found that
the equations of our consideration have the Painlev\'{e} property \ and the
analytic solution can be expressed in terms of a Right Painlev\'{e} Series.

\section{Acknowledgements}

AKH expresses grateful thanks to UGC (India), NFSC, Award No.
F1-17.1/201718/RGNF-2017-18-SC-ORI-39488 for financial support and Late
Prof. K.M.Tamizhmani for the discussions which AKH had with him which formed
the basis of this work. PGLL acknowledges the support of the National
Research Foundation of South Africa, the University of KwaZulu-Natal and the
Durban University of Technology.

\end{document}